\documentstyle[12pt,epsf]{article}

\def\half{{1\over 2}}
\def\ben{\begin{equation}}
\def\een{\end{equation}}
\def\bena{\begin{eqnarray}}
\def\eena{\end{eqnarray}}

\newcount\hour
\newcount\mhour
\newcount\minute
\newcount\tenminute
\newcount\mtenminute
\hour=\time \divide\hour by 60
\mhour=\hour \multiply\mhour by 60
\minute=\time \advance\minute by -\mhour \tenminute=\minute \divide\tenminute by 10 \mtenminute=\tenminute \multiply\mtenminute by 10 \advance\minute by -\mtenminute
\edef\timeofday{\the\hour:\the\tenminute\the\minute} \edef\today{%
\ifcase\number\month \or January \or February \or March \or April \or May \or June \or July \or August \or September \or October \or November \else December \fi
\ifcase\number\day \or 1st \or 2nd \or 3rd \or 4th \or 5th \or 6th \or 7th \or 8th \or 9th \or 10th \or 11th \or 12th \or 13th \or 14th \or 15th \or 16th \or 17th \or 18th \or 19th \or 20th \or 21st \or 22nd \or 23rd \or 24th \or 25th \or 26th \or 27th \or 28th \or 29th \or 30th \or 31st \fi
\number\year\quad}

\input amssym.def
\input amssym.tex

\newcommand {\nn}{\nonumber \\}

\newcommand {\ka}{\kappa}

\newcommand {\na}{\nabla}

\newcommand {\Del} {\Delta}

\newcommand {\third} {\frac{1}{3} }

\newcommand {\sqg} {\sqrt{g}}

\newcommand {\Lcal}{{\cal L}}

\newcommand {\pr} {{\quad .}}
\newcommand {\com} {{\quad ,}}
\newcommand {\q} {\quad}

\newcommand {\nl} {\newline}
\newcommand {\vs}[1] { \vspace*{#1 cm} } 

\newcommand {\NP} {Nucl.Phys.}
\newcommand {\PL} {Phys.Lett.}
\newcommand {\PR} {Phys.Rev.}

\newcommand {\JMP} {Jour.Math.Phys.}

\begin{document}

\hfuzz=100pt
\title{The Finiteness Requirement for Six-Dimensional Euclidean Einstein Gravity 
\footnote
{Nov. 1999,
hep-th/9911167,
LPTENS-99.48,AEI-1999-37}
}
\author{G.W.\ Gibbons
\footnote{On leave of absence from DAMTP, Cambridge University, Cambridge, UK.
E-mail address: G.W.Gibbons@damtp.cam.ac.uk, address from 1st Jan 2000 to 30th April 2000: Yukawa Institute for Theoretical Physics, Kyoto University Kyoto, 606-8502, Japan)}
\ and S.\ Ichinose 
\footnote{
On leave of absence from
Department of Physics, University of Shizuoka, Yada 52-1, Shizuoka 422-8526, Japan (Address after Nov.24, 1999). E-mail address:\ ichinose@u-shizuoka-ken.ac.jp 
}
\\
\\
\\ Laboratoire de Physique Th\'eorique de l'Ecole Normale Sup\'erieure \footnote{ Unit'e Mixte de Recherche du Centre National de la Recherche Scientifique et de l'Ecole Noemale Sup\'erieure} ,
\\ 24 Rue Lhomond,
\\ 75231 Paris Cedex 05, France
\\
\\ $\mbox{}^\ddag$	
Albert Einstein Institut
\\ Max-Planck-Institut f{\" u}r Gravitationspysik \\ Am M{\" u}hlenberg, Haus 5, D-14476 Golm, Germany 
}
\maketitle
\begin{abstract}
The finiteness requirement for Euclidean Einstein gravity is shown to be so stringent that only the flat metric is allowed. We examine counterterms
in 4D and 6D Ricci-flat manifolds
from general invariance arguments.
\end{abstract}

\vspace{1cm}
PACS NO\ :\ 04.60.-m, 
11.10.-z 
\nl
Key Words\ :\ Finiteness, Euclidean Einstein Gravity, Counterterm, Flat Metric, Lichnerowicz identity, Graphical Representation. 
\section{Introduction}

In a recent note by one of us\cite{Gib99} it was shown, by using an old result of Lichnerowicz,
that the
only positive definite (Riemannian, often called Euclidean)4-metrics for which the integrated two loop counterterms of pure Einstein gravity vanish are flat.
This is in contrast with Lorentzian 4-metrics, some of which may have all counterterms vanishing pointwise without being flat.

In fact in four dimensions,
given the vanishing of the Ricci tensor, the one loop counterterm is the local integrand in the Gauss-Bonnet expression for the Euler number and is proportional to the square of the Riemann curvature tensor. Thus, as long as the metric is positive definite, 
already at one loop, the counterterm can only vanish if the metric is flat. For a Lorentzian pp-wave metric however the square of the Riemann tensor necessarily vanishes. Note that although the integrand is locally a total derivative it is not the divergence of a covariant vector field and thus it cannot be disregarded as a counterterm. It definitely contributes around non-trivial backgrounds. However, in view of its topological nature, one might imagine absorbing this divergence by renormalizing a topological coupling constant. The new result at two loops then tells us that even disregarding the one loop divergence there are no non-flat two-loop finite metrics in pure gravity for which the integrated counterterm vanishes.
In this paper we shall discuss the situation in six dimensions at one-loop.

By contrast with the case of four dimensions, the one loop counterterm is not purely topological, although it contains a contribution proportional to the local integrand of the Gauss-Bonnet theorem (also called the Euler term), there are extra terms. Another difference is that the divergence identity of Lichnerowicz comes in at one loop rather than two loops as it does in four dimensions. Our final result is nevertheless similar to that in four dimensions: the integrated extra terms can only vanish in a flat background. However, as we shall see, 
this does not seem to preclude a possible cancellation of the extra terms against the Euler number. We will examine the exceptional case separately. 

Before embarking on the calculations, we wish to expand on some of the motivations for this work. In the past few years particular solutions
of the equations of the {\sl classical} equations of motion of gravity and super-gravity theories have been extensively used to investigate the {\sl quantum} properties of string theory and M-theory. The best known examples are so-called BPS solutions, i.e. those admitting Killing spinors. Because of the supersymmetry many, but not all BPS solutions
are believed to suffer no quantum corrections and so the properties of those classical solutions should persist at the quantum level. Typically the reasons for believing that quantum corrections vanish are so called "non-renormalization" theorems which are based on the pointwise vanishing of counterterms on these backgrounds. An example is provided by self-dual solutions of the Euclidean Einstein equations, considered as solutions of $N=1$ supergravity theories. Of course $N=1$ supergravity is not generally believed to be a consistent quantum theory of gravity, but is believed to be a consistent low-energy approximation to string theory. Thus one has some confidence that some of the properties of these classical solutions will persist in the full quantum theory.

Another example is provided by pp-waves of the Lorentzian vacuum Einstein equations. These certainly admit Killing spinors and are hence BPS, but they also possess another very striking property: all invariants formed from the curvature tensor vanish. In fact because of the structure of the curvature tensor, it seems likely that the pp-waves will be solutions of almost any set of covariant field equation (obtained possibly by taking the variation of some effective action) without a cosmological
term of the form
\ben
R_{ab}=S_{ab}(g_{ab}, R_{abcd}, R_{abcd;c}\dots) \label{Effective}, \een
where $S_{ab}$ is a trace free tensor constructed from the metric, the curvature tensor and its covariant derivatives. Physically pp-waves represent gravitational waves and from the above it would seem that we can be pretty confident that gravitational wave solutions of some effective action in the quantum
theory behave very much like
gravitational wave solutions in the classical theory. In particular we expect no modification of their properties as they propagate freely through empty spacetime\cite{GibbonsW, SD}. In the case of pp-waves we were not using any particular form of the effective action, just the vanishing of the invariants and the variational derivatives with respect to the metric.

Another, slightly more trivial example of metrics whose properties will be essentially unchanged, up to a scale, by quantum effects is provided by spaces of constant curvature, i.e. Anti-de-Sitter and de-Sitter spacetime. Substituting the expression $R_{abcd}=c(g_{ac}g_{bd}-g_{ad}g_{bc})$ into the effective equations of motion (\ref{Effective}) but where now $S_{ab}$ is no longer trace free, gives an equation for the constant $c$ which generically will have a number of real solutions. In fact the quantum corrections in this case will merely shift the radius form its classical value. Because of the very simple structure of their curvature tensors, Anti-de-Sitter space and de-Sitter space are examples of what one might call universal solutions. This concept is close to Bleecker's idea of "critical metrics" in Riemannian geometry \cite{bleecker, GibbonsT}. These are metrics which are critical points of {\sl any} diffeomorphism invariant action functional constructed from the metric and its derivatives. In that case he showed that critical metrics are homogenous spaces $M=G/H$ where $H$ acts irreducibly on the tangent space. 

It is clearly desirable to discover as many of these privileged classes of metrics as possible. They obviously do not exhaust all possibly relevant solutions of the effective equations of motion, but they are
ones whose properties we can be fairly confident of. In general we can only hope to find solutions of the effctive equations of motion in some sort of perturbation series whereas the solutions we are seeking are classically exact. Thus the search for them is analogous to the search for exact solutions in classical
general relativity but our criterion is more stringent. 

In the Lorentzian
case there seem to be a number of other examples in addition to those already mentioned, for instance pp-waves moving in Anti-de-Sitter spacetime \cite{siklos, ruback} are very likely "critical" or universal. In addition there exist some four-dimensional Lorentzian spacetimes of Petrov type N and III which have all invariants vanishing 
\footnote{Of course the vanishing of the invariants does not by itself mean that their variational derivatives vanish but it does mean that the integrals of the counterterms vanish identically.} 
\cite{Gib99}. In the Riemannian case, the set of critical metrics is from Bleecker's results, rather small. Thus one may relax it to demanding that integrals of some or all invariants vanish. In four dimensions at cubic order in the curvature tensor, i.e. at two loops, there is only one invariant which is not a total derivative available. It was shown that, among Ricci flat metrics, the integral can vanish only in the trivial flat case. In particular the integral is non-vanishing for self-dual spaces. Because only one invariant
is available in this case,
it must correspond to the local counterterm of any two loop non-finite theory of gravity, such as Einstein gravity. We deduce
that the quantum finiteness of self-dual spaces depends in an essential way on embedding
them in a supersymmetric theory.

In this paper we shall be concerned with the six-dimensional case. As indicated above this is more complicated. It is nevertheless of considerable physical interest. Firstly, in superstring compactifications one has a six-dimensional Ricci flat internal space which is usually taken to be a Calabi-Yau space. This admits covariantly constant spinors and is a supersymmetric solution of the low-energy supergravity approximation to string theory. Secondly such spaces play a role in the
theory of the M-5-brane ( see e.g. \cite{tseytlin}) where one considers the conformal counterterms associated with the (2,0) tensor multiplet.

The plan of the paper is as follows. In section 2 we shall, for the convenience of the reader, recall the essential deatils of the four-dimensional case. In section 3 we discuss the six-dimensional case. Section 4 contains a discussion and a conclusion.

\section{4 Dimensional Euclidean Gravity} Let us first recall the case of four dimensional pure Einstein gravity with a positive definite (Euclidean) metric \cite{Gib99}. The action is 
\bena
\Lcal=\frac{1}{\ka}\sqg R\com
\label{FD1}
\eena
where $\ka$ and $R$ are the gravitational constant and the Riemann scalar curvature.
Their physical dimensions
are $[\ka]=\mbox{\{M(ass)\}}^{-2}$ and $[R]=\mbox{M}^{2}$ respectively.
In perturbative quantum gravity, the original Lagrangian must be shifted as
$\Lcal+\Del\Lcal^{\mbox{1-loop}}+\Del\Lcal^{\mbox{2-loop}}+\cdots$, in order to subtract ultraviolet divergences. 
The most general 1-loop counter Lagrangian, not neglecting the total derivative terms
\footnote{
In many conventional treatments of counterterms, total derivatives and the Euler term are ignored\cite{tHV}. Here we keep them in order to know how the requirement of finiteness constrains the global and asymptotic behavior of the spacetime manifold. }
, may be written as
\bena
\Del\Lcal^{\mbox{1-loop}}=a
\na^2R+b_1 R^2 +b_2 R_{ab}R^{ab}+b_3 R_{abcd}R^{abcd}\pr \label{FD2}
\eena
The 4 terms above are locally independent and the complete list of
general invariants with the dimension of M$^{4}$. \footnote{
If we identify 1-loop counterterm with the trace anomaly, then
the possible terms reduce to the following three \cite{Duff77,DS93,SI98}\ :\
$E$(Euler term),$C$(Conformal invariant) and $\na^2 R$(trivial term).
}
Here we introduce a convenient graphical representation\cite{SI95} for these invariants which facilitates algebraic manipulations, especially for the extension to the higher orders
and to the higher dimension.
\bena
\na^2R=
\begin{array}{c}
\mbox{
{\epsfysize=8mm\epsfbox{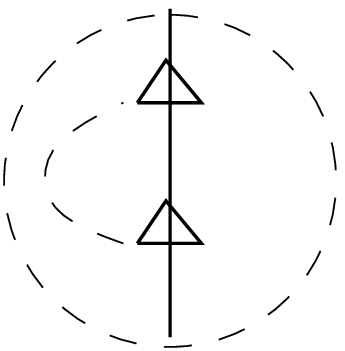}} 
}
\end{array}\com\q
R^2=
\begin{array}{c}
\mbox{ {\epsfysize=8mm\epsfbox{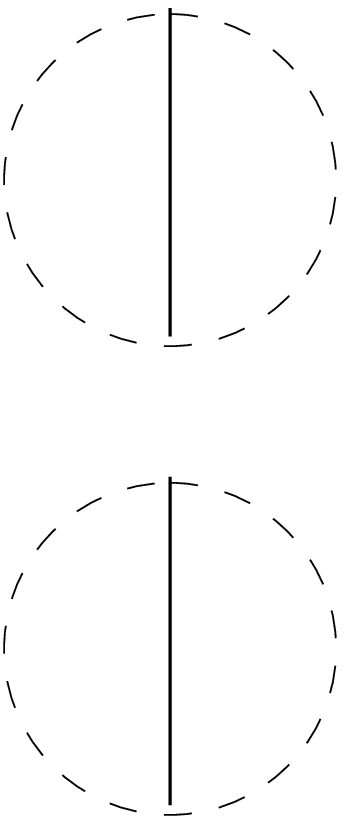}} } 
\end{array}
\com\nn
R_{ab}R^{ab}=
\begin{array}{c}
\mbox{ {\epsfysize=8mm\epsfbox{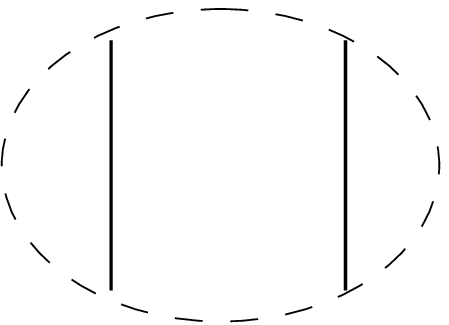}} } 
\end{array}\com\q
R_{abcd}R^{abcd}=
\begin{array}{c}
\mbox{ {\epsfysize=8mm\epsfbox{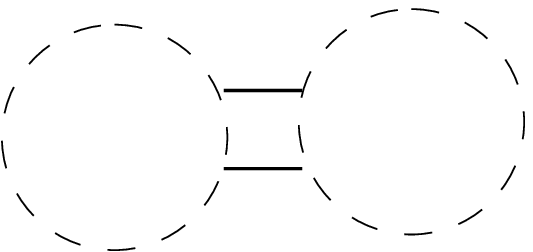}} } 
\end{array}
\pr
\label{FD2B}
\eena
Among the 4 invariants above, two special combinations are well known:\ the Euler term ($E$) and the conformal invariant ($C$). 
\bena
\mbox{Euler term}\q E=
R^2-4R_{ab}R^{ab}+R_{abcd}R^{abcd}\com\nn \mbox{Conformal invariant}\q C=
\third R^2-2 R_{ab}R^{ab}+R_{abcd}R^{abcd}\pr \label{FD3}
\eena
Now we consider Ricci flat (RF) manifolds: 
\bena
R_{ab}= 0.
\label{FD4}
\eena
The quantities above reduce to
\bena
\Del\Lcal^{\mbox{1-loop}}|_{RF}=b_3
\begin{array}{c}
\mbox{ {\epsfysize=8mm\epsfbox{R4R4.eps}} } 
\end{array}
\com\q
E|_{RF}=C|_{RF}=
\begin{array}{c}
\mbox{ {\epsfysize=8mm\epsfbox{R4R4.eps}} } 
\end{array}
\pr
\label{FD5}
\eena
Thus the requirement of {\it Ricci flat (on-shell) finiteness} leads to (assuming $b_3\neq 0$)
\bena
\int\sqg
\begin{array}{c}
\mbox{ {\epsfysize=8mm\epsfbox{R4R4.eps}} } 
\end{array}
=\int\sqg R_{abcd}R^{abcd}=0\pr
\label{FD6}
\eena
This means, in the Euclidean metric, that the geometry is {\it locally flat}:
\bena
R_{abcd}=0\pr
\label{FD7}
\eena
Consequently this says $E$ and $C$ locally vanish. In particular the Euler number vanishes:\ $\int\sqg E|_{RF}=0$.

One might wonder whether this result is an accidental due to the simplicity of the 1-loop terms or the special nature of the 4 dimensions one 1-loop counterterm, which is proportional to
the Euler term for a Ricci flat manifold. Therefore let us examine the 2-loop terms. The most general form of the 2-loop counterterm, including total derivatives, can be expressed as
some linear combination of the following 15 terms. 
\bena
\frac{1}{\ka}\Del\Lcal^{\mbox{2-loop}}=
\sum^5_{i=1}x_iP_i+wA_1+\sum^4_{i=1}y_iO_i+\sum^4_{i=1}z_iT_i +vS\com
\label{FD8}
\eena
where $x_i,w,y_i,z_i,v$ are some constants. $P_1\sim P_5; A_1;O_1\sim O_4;T_1\sim T_4; S$
express $M^6$-invariants and are defined in \cite{II97} with their graphs. (See eq.(24),eq.(43) and Figs.44-48 of this reference). On a Ricci flat manifold,$\Del\Lcal^{\mbox{2-loop}}$(\ref{FD8}) reduces to,
\bena
\frac{1}{\ka}\Del\Lcal^{\mbox{2-loop}}|_{RF}= wA_1+y_3O_3+z_3T_3\com
\label{FD9}
\eena
where
\bena
A_1=
\begin{array}{c}
\mbox{ {\epsfysize=8mm\epsfbox{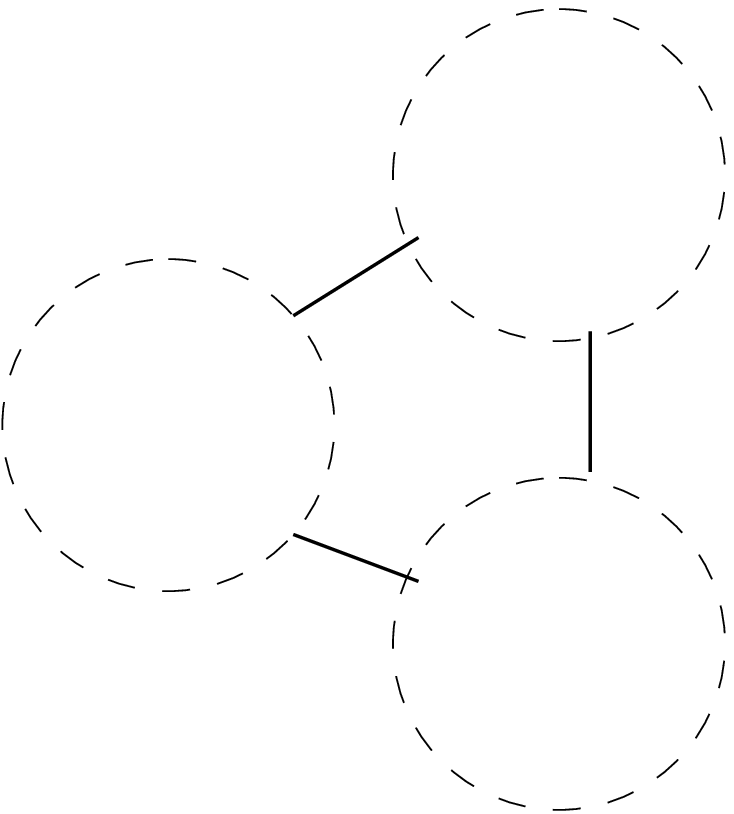}} } 
\end{array}
\com\q O_3=
\begin{array}{c}
\mbox{ {\epsfysize=8mm\epsfbox{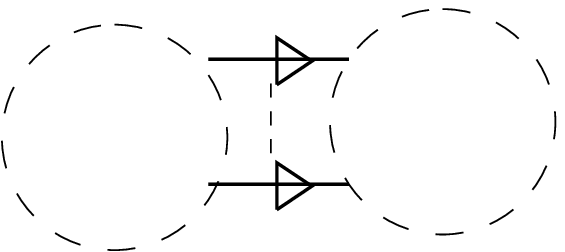}} } 
\end{array}
\com\q T_3=
\begin{array}{c}
\mbox{ {\epsfysize=8mm\epsfbox{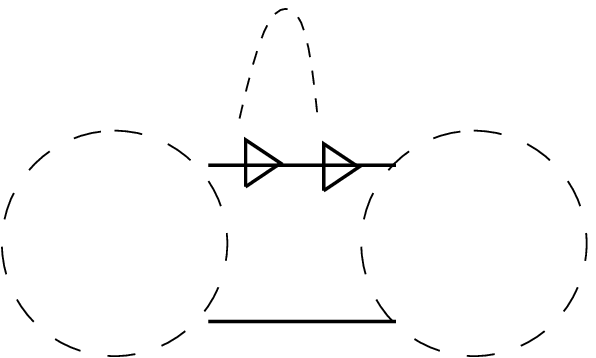}} } 
\end{array}
\pr
\label{FD10}
\eena
In conventional index notation,
$A_1=R_{abcd}R^{dc}_{~~ef}R^{feba},
O_3=\na_eR_{abcd}\cdot\na^eR^{abcd},
T_3=R_{abcd}\na^2R^{abcd}$.
\footnote{
$A_1=-I_3$ where $I_3$ is that one used in the original paper \cite{Gib99}. $O_3$ and $T_3$ look like ``descendants'' of 1-loop on-shell counterterm($R_{abcd}R^{abcd}$). }
Here we note $O_3+T_3$ is a total derivative. 
\bena
O_3+T_3=\na_a K^a\com\q
K^a=
\begin{array}{c}
\mbox{ {\epsfysize=16mm\epsfbox{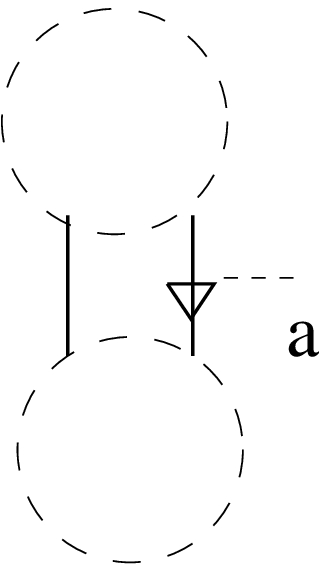}} } 
\end{array}
\com
\label{FD11}
\eena
where $K^f=R_{abcd}\na^fR^{abcd}$.

Before imposing 2-loop finiteness, we rewrite $A_1$ in (\ref{FD9}) using the Lichnerowicz identity\cite{Lich77} on a 4 dim Ricci flat manifold:
\bena
\na^2(
\begin{array}{c}
\mbox{ {\epsfysize=8mm\epsfbox{R4R4.eps}} } 
\end{array}
)=6A_1+2O_3\pr
\label{FD12}
\eena
(See Appendix.) Finally (\ref{FD9}) is written as 
\bena
\frac{1}{\ka}\Del\Lcal^{\mbox{2-loop}}|_{RF}= \frac{w}{6}\na^2
\begin{array}{c}
\mbox{ {\epsfysize=8mm\epsfbox{R4R4.eps}} } 
\end{array}
+(-\frac{w}{3}+y_3-z_3)O_3+z_3\na_a K^a\pr \label{FD13}
\eena
We consider a \lq\lq regular" manifold on which the total integral of total derivatives vanish.
Then the requirement of 2-loop finiteness reduces to, assuming $-\frac{w}{3}+y_3-z_3\neq 0$,
\bena
\int\sqg O_3=\int\sqg (\na_e R_{abcd})^2=0\pr \label{FD14}
\eena
This means, in the case of a Euclidean metric, 
\bena
\na_e R_{abcd}=0\pr
\label{FD15}
\eena
Using this result, we obtain
\bena
\na_e
\begin{array}{c}
\mbox{ {\epsfysize=8mm\epsfbox{R4R4.eps}} } 
\end{array}
=2R^{abcd}\na_e R_{abcd}=0\pr
\label{FD16}
\eena
In the asymptotically locally flat (ALF) or asymptotically locally Euclidean (ALE) case, (\ref{FD16}) imply
\bena
\begin{array}{c}
\mbox{ {\epsfysize=8mm\epsfbox{R4R4.eps}} } 
\end{array}
=0\com
\label{FD17}
\eena
which means again (\ref{FD7}).

We claim that to require finiteness
is to require a locally flat metric (\ref{FD7}) in 4 dim Euclidean pure Einstein gravity. We have shown this for 1-loop and 2-loop orders. It would be quite interesting if one could prove the same result for higher orders. The key point for such an extension would be a generalization of the Lichnerowicz identity to general invariants with higher dimensions as shown in Fig.1. 
\begin{figure}
3-loop(8 dim):\
{\epsfysize=20mm\epsfbox{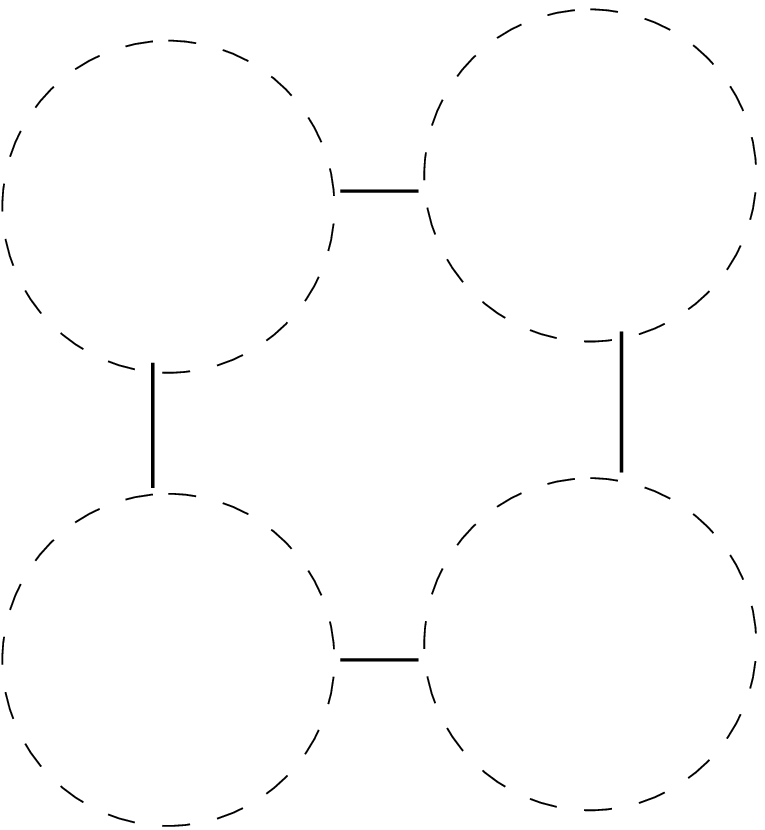}} 
\ ,\
4-loop(10 dim):
{\epsfysize=20mm\epsfbox{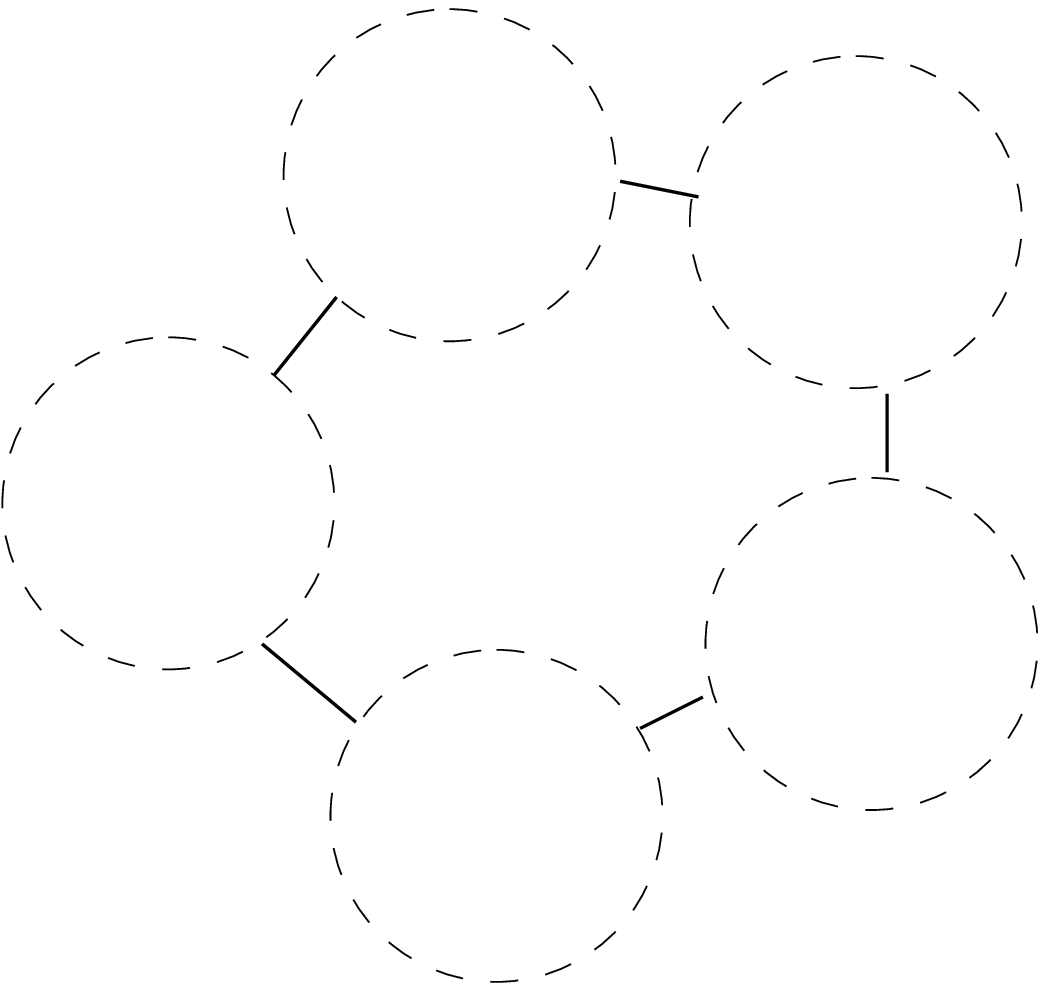}} 
\ ,\ $\cdots$
%
%
\begin{center}
Fig.1\
Key graphs for the higher orders (or higher dimensions) generalization of the present result.
\end{center}
\end{figure}

\section{6 Dimensional Euclidean Gravity} Besides the higher loop order situation, the higher dimensional extension is also interesting. Let us consider 6 dim pure Einstein gravity with a Euclidean metric. 
\bena
\Lcal_6=\frac{1}{\ka_6}\sqg R\com\q
[\ka_6]=\mbox{M}^{-4}\com\q [R]=\mbox{M}^{2}\pr \label{SD1}
\eena
The most general 1-loop counter Lagrangian, including total derivatives, may be written as
\bena
\Del\Lcal_6^{\mbox{1-loop}}=
\sum^6_{i=1}x_iP_i+w_1A_1+w_2B_1+\sum^4_{i=1}y_iO_i+\sum^4_{i=1}z_iT_i +vS \pr
\label{SD2}
\eena
The 17 terms above are defined in eq.(24) of Ref.\cite{II97}. Note that above expression
slightly differs from (\ref{FD8}) in that $P_6$ and $B_1$ appear here. \footnote{
Let $n$ and $d$ even and $d<n$, then there generally appears some relations,
among $M^n$-invariants in $d$-dim space, which are special to the space dimension $d$. Terms $P_6$ and $B_1$ are not in  eq.(\ref{FD8}), where $n=6$ and $d=4$,  due to the special relations. }
On a Ricci flat manifold,
(\ref{SD2}) reduces to
\bena
\Del\Lcal_6^{\mbox{1-loop}}|_{RF}=
w_1A_1+w_2B_1+y_3O_3+z_3T_3\com
\label{SD3}
\eena
where $A_1,O_3,T_3$ are given in (\ref{FD10}), and 
\bena
B_1=
\begin{array}{c}
\mbox{ {\epsfysize=8mm\epsfbox{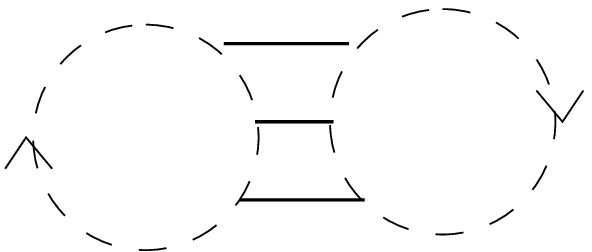}} } 
\end{array}
\pr
\label{SD4}
\eena
($B_1=R_{abcd}R^{b~~c}_{~ef}R^{eadf}$.)
There are three Conformal (Weyl)invariants $C_1,C_2,C_3$ and one Euler term $E$ . They are, on a Ricci flat manifold, 
\bena
C_1|_{RF}=A_1\com\q C_2|_{RF}=B_1\com\q C_3|_{RF}=-5T_3\com E|_{RF}=4A_1-8B_1\pr\label{SD5}
\eena
(cf.(\ref{FD5})). Using the following three relations: 
\begin{description}

\item{i)} the Lichnerowicz identity\cite{Lich77} on a six dimensional Ricci flat manifold is,
\bena
\na^2 (
\begin{array}{c}
\mbox{ {\epsfysize=8mm\epsfbox{R4R4.eps}} } 
\end{array}
)=2A_1+8B_1+2O_3\com
\label{SD6}
\eena
(See Appendix, cf (\ref{FD12})),

\item{ii)} the Euler term relation in (\ref{SD5}), and

\item{iii)} the relation (\ref{FD11}),

\end{description}
\noindent eq. (\ref{SD3}) can be rewritten as 
\bena
\Del\Lcal_6^{\mbox{1-loop}}|_{RF}=
\frac{1}{12}(2w_1+w_2)\na^2
\begin{array}{c}
\mbox{ {\epsfysize=8mm\epsfbox{R4R4.eps}} } 
\end{array}
+\frac{1}{24}(4w_1-w_2)E|_{RF}\nn
+\{ -\frac{1}{6}(2w_1+w_2)+y_3-z_3\} O_3+z_3\na_a K^a\pr \label{SD7}
\eena
(cf. (\ref{FD13})). Now we require 1-loop finiteness: 
\bena
\int d^6x\sqg\Del\Lcal_6^{\mbox{1-loop}}|_{RF}=0\pr \label{SD8}
\eena
We shall assume that the boundary terms of genuinely covariant divergence identities vanish. However even if we do so, we cannot assume in general that the integral of the Gauss-Bonnet integrand vanishes.
We notice here a delicate thing appers: an analogous one appeared at 1-loop for 4 dim gravity.
In order to clearly look at the things which are independent of the Euler number contribution, we treat the following two cases separately. 
\bena
\mbox{Case A}\q:\q \int d^6x\sqg E|_{RF}=0\com\label{SD9A}\\ \mbox{Case B}\q:\q \int d^6x\sqg E|_{RF}\neq 0\pr \label{SD9B}
\eena
See footnote below \footnote{
In the 4 dim case(see section 2), the vanishing of the Euler number is deduced
from the 1-loop finiteness. Vanishing of the volume integral of the boundary terms is required at 2-loop level.
}.
\medskip

Case A\nl

In this case, with the assumption:\
$-\frac{1}{6}(2w_1+w_2)+y_3-z_3\neq 0$, the 1-loop finiteness requirement (\ref{SD8}) again reduces to (\ref{FD14}). Because the results of (\ref{FD15}-\ref{FD17}) hold true for 6 dim, we conclude that the requirement of (1-loop) finiteness again means (\ref{FD7}) that the metric is locally flat. 

If our manifold is compact then (\ref{SD9A}) implies that the Euler characteristic
vanishes. In four dimensions the Ricci-flat condition would then force the metric to be flat. In six dimensions closed Ricci flat, non-flat manifolds may have vanishing Euler number. The product of a two dimensional torus with a K3 surface is an example. If the six-manifold is not closed then there will in general be boundary contributions to the Gauss-Bonnet formula and (\ref{SD9A}) does not necessarily imply the vanishing of the Euler number. \nl
\nl

Case B\nl

In Case B, the Euler number contribution cancels against the remaining contribution.
\bena
+\frac{1}{24}(4w_1-w_2)\int d^6x\sqg E|_{RF} +\{ -\frac{1}{6}(2w_1+w_2)+y_3-z_3\} \int d^6x\sqg O_3=0\pr \label{SD10}
\eena
First we should note that the two coefficients, in front of above two terms, are determined by the ultraviolet structure of the quantum gravity. While
the integral of the the Gauss-Bonnet term, $\int d^6x\sqg E|_{RF}$, is determined by the topology of the background space manifold. Therefore, assuming
$
4w_1-w_2\neq 0,\
-\frac{1}{6}(2w_1+w_2)+y_3-z_3\neq 0
$,
the above equation requires the volume integral of a local quantity, $\int d^6x\sqg O_3$, is determined only by the local quantum structure and the global structure of the background manifold. If such a metric consistently exists, it must be a very special one ( which describes something like a ``boundary'' theory ).

\section{Discussion and Conclusion}
The coefficients of the counterterms in 6 dim pure Einstein gravity (with Lorentzian metric) was obtained by P.van Nieuwenhuizen\cite{Nieu77} and by\break R. Critchley\cite{Cri78}. They focused on $A_1$ ( not on $O_3$ as in the present analysis). The Ricci flat counter Lagrangian (\ref{SD7}) can also be expressed as

\bena
\Del\Lcal_6^{\mbox{1-loop}}|_{RF}=
(w_1+\half w_2-3y_3+3z_3)A_1 \nn
+\mbox{Euler-term}
+\mbox{Total-derivative-terms}\pr
\label{COM1}
\eena
They obtained the coefficient in front of $A_1$ as $\frac{9}{4\pi^3 15120}$. Note that this coefficient is proportional to the previous one in front of $O_3$ and it shows the non-zero assumption taken there holds true. In their
derivation, in effect, the Lichnerowicz identity was used. In the eq.(63) of Ref.\cite{Cri78}, the top equation corresponds to the identity (\ref{SD6}), the middle one to (\ref{FD11}), and the bottom to the first equation of (\ref{SD5}). ($x$ and $y$, in their notation, correspond to $-A_1$ and $-B_1$ respectively.)

The approach to the finiteness of the quatum gravity taken in \cite{Gib99} and the present paper should be distinguished from the ordinary one taken so far. Ordinarily, beginning from the 't Hooft and Veltman's analysis\cite{tHV}, the focus is mainly on the cancellation between coefficients of counterterms, where care is not taken so much for
the background (metric) field except that it satisfies the field equation. In the ordinary approach, the supergravity
theories realize the finiteness requirement to some extent. The 1-loop cancellation in some theories is reviewed in \cite{Duff94} from the view of the Weyl anomaly.
Information about the counterterms for scalars, spinors and vectors is given in \cite{cri78}. 
Quite recently the conformal anomaly in the free D=6 superconformal (2,0) 
tensor multiplet theory on the curved background has been computed
\cite{tseytlin}.
(They compare the result with that of AdS/CFT. Some discrepancy ( for the
Euler term) appears and they say the free tensor multiplet 
anomaly does not
vanish on the Ricci flat manifold.)
The present approach, by contrast, 
focuses on the background metric field itself
rather than on the coefficients.
Instead of seeking the
cancellation among coefficients,
the constraints on the metric field which are induced from the finiteness requirement are examined. The standpoint is that the quantum structure can constrain the effective background field. Our observation is that the finiteness requirement in the present approach is so stringent,
in Euclidean case, that the flat space is only allowed (except a special case).

\vs 1
\begin{flushleft}
{\bf Acknowledgment}
\end{flushleft}
One of us (S.I.) thanks the hospitality at the Albert-Einstein-Institut where this work has been finished. He also thanks the Japanese Ministry of Education for the partial financial support. The other (G.W.G.) would like to thank the \'Ecole Normale and the Yukawa Institute for support and hospitality during the preparation of this paper.

\vspace{2cm}
{\LARGE Appendix\ \ Lichnerowicz identity}\cite{Lich77}\nl \nl

The following identity holds for {\it any} dimension. 

\bena
\na^2
\begin{array}{c}
\mbox{ {\epsfysize=8mm\epsfbox{R4R4.eps}} } 
\end{array}
=K+2
\begin{array}{c}
\mbox{ {\epsfysize=8mm\epsfbox{DR4DR4.eps}} } 
\end{array}
\com\nn
K=8
\begin{array}{c}
\mbox{ {\epsfysize=8mm\epsfbox{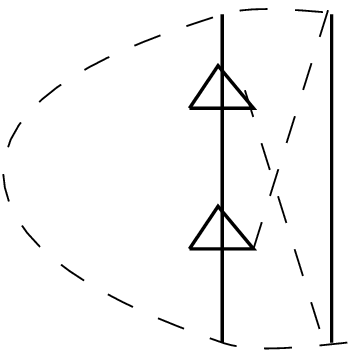}} } 
\end{array}
-4
\begin{array}{c}
\mbox{ {\epsfysize=8mm\epsfbox{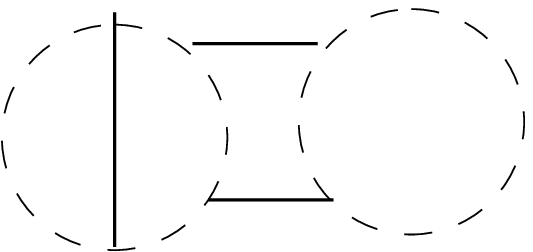}} } 
\end{array}
\nn
+2
\begin{array}{c}
\mbox{ {\epsfysize=8mm\epsfbox{RRR.eps}} } 
\end{array}
+8
\begin{array}{c}
\mbox{ {\epsfysize=8mm\epsfbox{RRRb.eps}} } 
\end{array}
\pr
\label{Lich1}
\eena
On a Ricci flat manifold, $K$ reduces to 
\bena
K|_{RF}=
2
\begin{array}{c}
\mbox{ {\epsfysize=8mm\epsfbox{RRR.eps}} } 
\end{array}
+8
\begin{array}{c}
\mbox{ {\epsfysize=8mm\epsfbox{RRRb.eps}} } 
\end{array}
\pr
\label{Lich2}
\eena
For the space dimension less than 6, the above one further reduces to
\bena
K|_{RF}=
6
\begin{array}{c}
\mbox{ {\epsfysize=8mm\epsfbox{RRR.eps}} } 
\end{array}
\pr
\label{Lich3}
\eena

\vspace{2cm}

\end{document}